\documentstyle[preprint,aps]{revtex}

\begin{document}
\draft

\title{Spin-Charge Separation, Anomalous Scaling and the Coherence of Hopping
       in exactly solved Two Chain Models}

\author{Nic Shannon, Yanmin Li and Nicholas d'Ambrumenil}

\address{
Department of Physics, 
University of Warwick,
Coventry,
England\\}

\date{\today}
\maketitle

\begin{abstract}
The coherence of transport between 
two one-dimensional interacting Fermi liquids, coupled by single particle 
hopping and interchain interaction, is 
examined in the 
context of two exactly soluble models. 
It is found that the coherence of the inter-chain hopping 
depends on the interplay 
between inter-chain hopping and inter-chain interaction terms, 
and not simply 
on the ground state spectral properties of an isolated chain.  
Specifically, the splitting of levels in associated with 
interchain hopping in a $g_4$ soluble model is found to be enhanced by 
the introduction of interchain interaction.  
It is also shown that, for an exactly solvable model with both 
$g_2$ and $g_4$ interactions, coherent 
interchain hopping coexists with anomalous scaling and non-Fermi 
liquid behavior in the chain direction.

\end{abstract}
\narrowtext

\section{Introduction}

Although the physics of an ideal one-dimensional Fermi gas has been widely 
and profitably studied for more than twenty years \cite{emery,voit},
our understanding of real low-dimensional materials is still
limited by the lack of a clear picture of how the crossover in electronic
properties between 
strictly one-dimensional and three-dimensional limits occurs.
While it is clear that a one-dimensional interacting Fermi gas 
{\it cannot} be a Landau Fermi Liquid, the possibility of finding in higher 
dimensions a ground state 
analogous to the non-Fermi Liquid (NFL) ground states found
in one dimension (in particular the Luttinger 
Liquid (LL), characterized by different velocities for spin and charge 
excitations, 
and an anomalous exponent $\alpha$ controlling spectral properties) is still 
a subject of much debate.  Interest in
these issues has been stimulated by the unusual  
properties found in high temperature superconductors and 
the wide variety of other
material in which the electrons are believed to be both strongly 
correlated and confined to move in low dimension.

Theoretical consideration of two dimensional interacting Fermi liquids 
with fully two 
dimensional Fermi surfaces by a number of authors and techniques 
\cite{2Dboson,other} 
has consistently found that the Landau Fermi Liquid state {\it is}
stable, excepting cases of extremely long range
interaction which are of questionable physical relevance.  Coupled chain 
models (generally comprising a large number of identical one dimensional 
interacting Fermi liquids coupled by weak interchain hopping) have also  
attracted much attention 
\cite{finkelstein,koepitz,nersesyan1,nersesyan2,giamarchi,orignac,carol,noack,bourbonnais1,bourbonnais2,wen}.  
These  are 
directly applicable to the large class of quasi one-dimensional conductors 
which are held to be strongly correlated (for example the organic spin 
density
wave system $TTF-TCNQ$).  

The effects which might be expected to modify or destroy the 
one-dimensional properties of an isolated chain are interchain hopping 
and interchain interaction.  These are usually 
introduced as a single particle hopping between neighboring chains, and a 
density-density interaction between pairs of chains, with the 
possibility of backward scattering or umklapp terms in the interaction 
generally being neglected.  While the problem of chains coupled by 
interaction alone is soluble by the technique of bosonization 
\cite{wen,tomonaga,haldane}, interchain hopping spoils the special current
conservation laws (Ward identities) which make the solution of the
problem by bosonization techniques possible. 
In this regard it is analogous to the 
one-dimensional backward scattering problem \cite{luther}.

Scaling theories (perturbative renormalization group 
techniques) have been applied most intensively to the problem of two 
chains coupled by interchain hopping and interaction 
\cite{nersesyan2,bourbonnais1,orignac}, and have given 
considerable insight into the 
ground state instabilities (dominant fluctuations) found for different 
combinations of 
interchain and intrachain interaction.  These models have also been 
studied numerically \cite{carol,noack}.  The simplest scaling arguments 
\cite{bourbonnais1,wen} suggest that for spinless fermions, there is 
a critical value ($\alpha = \frac{1}{2}$ for spinless fermions) of the 
anomalous exponent associated with the one-dimensional Luttinger liquid 
ground state, above which weak hopping is an 
irrelevant perturbation.  Correlation of electrons on a single chain can 
then lead to their confinement within it.  This value of $\alpha$ can only 
be regarded as placing a bound on the stability of the one-dimensional
Luttinger liquid 
state.  It {\it does not} imply that systems formed form chains with 
$0 < \alpha < \frac{1}{2}$ automatically flow to a higher dimensional 
fixed point 
(i.e. Fermi Liquid); where interchain hopping is a relevant perturbation
it is not possible to determine the ground state properties of the 
coupled chain system from scaling arguments alone.
In fact the values of $\alpha$ considered applicable to real materials
are often less than 1/2.  For example, it is known that for a 
one-dimensional Hubbard chain  $\alpha \leq \frac{1}{8}$ in the limit 
$U \rightarrow \infty$.

The debate over the role played by interchain/interplane hopping 
in the stability of NFL states was extended by Anderson's suggestion 
\cite{anderson1} 
that weak single particle hopping between Luttinger Liquids 
cannot generate band structure in the direction of the hopping, even in 
cases 
where this hopping is relevant in the Renormalization Group Sense.  Instead
there is ``confinement by decoherence'': the 
system retains its non-Fermi Liquid (NFL) character in one (or two) 
directions, 
and suffers only diffusive motion of electrons perpendicular to this 
direction \cite{clarke,csa2,csa3}.   Since this method of confinement 
need not suppress 
Josephson Coupling (pair hopping) between NFL's, a set of weakly 
coupled NFL planes can become a three-dimensional (anisotropic) 
superconductor,
with c-axis transport properties matching those of the cuprates 
\cite{anderson2}.

Following this suggestion, the question of whether arbitrarily 
weak interchain hopping $t_{\perp}$ can cause a splitting of energy 
levels (analogous to the formation of binding and antibinding bands 
in the free electron case) in interacting two chain models has come 
to be seen as a useful first step towards understanding whether 
coherent transport is possible in the physically relevant N-chain 
case.  Two recent numerical studies \cite{noack,capponi} 
have found evidence of such a splitting in finite size 
systems.  In the case of the spinless two chain  
model studied by Capponi {\it et al.} \cite{capponi}, the splitting 
$\Delta E$ due to 
interchain hopping may be directly observed in the electronic 
spectral function, and decreases with increasing 
$\alpha$ from the free electron value of $2t_{\perp}$ (for 
$\alpha = 0$).  Finite size scaling of the results for $\Delta 
E(L)/2t_{\perp}$ 
as a function of $\alpha$, shows a finite but steadily decreasing 
splitting for all $\alpha < 0.3$, compatible with the results of 
scaling arguments. 

It is clear that, as the couplings between an array of chains embedded
in two or three dimensions become stronger, the system becomes more
nearly isotropic, and must eventually cross over to the two (three)
dimensional limit in which the Fermi liquid description may be applied.
This issue is complicated by the possibility of a crossover to an
ordered two- or three-dimensional state, stabilized by many particle
exchange between chains (as has been found in RG studies
of two chain models \cite{nersesyan2,bourbonnais1}).
The nature of this crossover was 
addressed by Anderson \cite{anderson1} and
recently by Tsvelik \cite{tsvelik1}, who 
identified the possibility of coherent
transport between chains with the existence of a well-defined pole in
the single particle Green's function.

In this paper we establish what may be 
said about the coherence of hopping between chains in two soluble 
extensions of one-dimensional models: a special case of the general 
two chain model displaying both spin charge separation and anomalous 
scaling, and an extension of the single branch spinful two 
chain model originally studied by Fabrizio and Tsvelik 
\cite{fab1,fab2,tsvelik2} to include interchain interaction.  
We will compare our 
results with existing numerical and perturbative analyses.  We begin by 
briefly reviewing Anderson's criterion for coherence 
and its application 
to coupled free electron chains.

\section{Criterion for Coherence}

The arguments about coherence presented by Clarke, Strong and Anderson 
(CSA) are rooted in an 
analogy between the problem of hopping between Luttinger Liquids and that 
of a two level system (TLS) coupled to a dissipative ohmic background 
\cite{chakravarty1,chakravarty2}.  In the light of this analogy, 
these authors propose 
that the criterion for coherence should be the probability of the system 
returning to a specially prepared state during its intermediate time 
evolution after the sudden introduction of hopping between chains at time 
$t=0$.  
The state in which the system is prepared is an eigenstate of the original 
Hamiltonian for $t < 0$ in which one of the two chains carries additional 
charge.  
This corresponds to the TLS having a non-zero probability of being 
found in one of its two states.  A fully coherent (e.g. free electron) 
system will evolve away from its original state, but return to it over a 
time scale set by the size of the hopping matrix element.  An incoherent 
system will never return to its original state - in the language of the 
dissipative TLS, the probability of it being found in either state must 
relax to zero.

We may formally consider the probability $P(t)$ of the system returning to 
its 
initial state

\begin {eqnarray}
P(t) & = & |A(t)|^2\nonumber\\
A(t) & = & <\psi_0 |e^{iH^{\prime}t} e^{-iHt}|\psi_0>\nonumber\\
H^{\prime} & = & H + H_{0 \perp}\nonumber\\
H|\psi_0> & = & E_0 |\psi_0>.
\label{eq:Poft}
\end{eqnarray}

\noindent
where $|\psi_0> $ is a specially prepared eigenstate of $ H $ with 
additional charge $\Delta N$ on one of the chains, and $H_{0\perp}$ is an 
interchain hopping term, generally taken to be of the form

\begin{eqnarray}
H_{0\perp} &=&  -t_{\perp} \sum_{k, \sigma} 
           [ c^{(1) \dagger}_{\sigma}(k) c^{(2)}_{\sigma}(k)
           + c^{(2) \dagger}_{\sigma}(k) c^{(1)}_{\sigma}(k) ] .
\end{eqnarray}
For free electron chains we have

\begin{equation}
H=H_{0 \parallel} = \sum_{k, \sigma} \epsilon (k) 
                    [ c^{(1) \dagger}_{\sigma}(k) c^{(1)}_{\sigma}(k)
                    + c^{(2) \dagger}_{\sigma}(k) c^{(2)}_{\sigma}(k) ], 
\end{equation}
and the overlap $A(t)$ may be evaluated exactly; the 
system can be treated as a product of two level (chain) systems for each 
wave number $k$ along the chains, and the Hamiltonian for each wave 
number (and spin index) separately diagonalized. The occupation of
states labeled by
$k$, which are initially occupied on one chain but not on the 
other, then oscillates between chains with frequency  $2t_{\perp}$, 
so that in the presence in the initial state of extra charge $\Delta N$ on 
one chain, we find \cite{correction}
\begin{equation}
P(t) = \cos^{2\Delta N} (t_{\perp}t).
\end{equation}

More generally, we may expand $|\psi_0>$ in the eigenstates of $H^{\prime}$
\begin {eqnarray}
|\psi_0> & = & \sum_n c_n |\phi_n>\nonumber\\
H^{\prime}|\phi_n> & = & \epsilon_n|\phi_n>\nonumber\\
A(t) & = & \sum_{n,m}c^*_m c_n
        <\phi_m|e^{-iH^{\prime}t}|\phi_n>e^{iE_0}\nonumber\\
     & = & \sum_n |c_n|^2e^{i(E_0-E_n)t}
\end {eqnarray}
Our model is not dissipative, and the quantity $P(t)$ possesses full time 
reversal symmetry, but the overlap $A(t)$ can display damping due to the 
interference between modes of different frequency $E_n$.  In finite size 
systems, where one might expect $P(t)$ eventually to return to one, damping 
can still be observed over all physical time-scales \cite{baldwin}. 

In order to extend their macroscopic analysis to correlated systems, 
CSA study the rate of decay of the system from its original 
state after the introduction of hopping 
using Fermi's Golden rule.  They claim that an anomalous exponent 
$\alpha > \frac{1}{4}$ is sufficient to impose a lower bound $t_{\perp}^{c}$
on the value of $t_{\perp}$ needed to generate coherent interchain hopping, 
so for Luttinger Liquids with $\alpha$ exceeding $\frac{1}{4}$, 
arbitrarily weak interliquid hopping {\it cannot} generate band structure
\cite{clarke,csa2,csa3}.
 
Mila and Poilblanc have calculated the Fourier transform of the overlap 
$A(t)$ 
numerically for a number of $2 \times L$ t-J Ladders \cite{frederic}, and 
find 
that the coherence demonstrated by this quantity in the limit 
$t_{\perp} \rightarrow 0$ is not simply determined by the anomalous 
scaling 
parameter found in the spectral function for an isolated chain, as 
suggested in 
CSA, but is considerably enhanced near integrable 
points of the single chain t-J plane.  They suggest that this may be 
understood 
in terms of the level statistics of the problem; for an integrable system 
the 
eigenstates with additional charge $\Delta N$ are highly degenerate.  
Degenerate perturbation theory must therefore be applied tending to yield a 
splitting in energy levels linear in $t_{\perp}$ and associated 
coherence.  One could also say that level repulsion in a 
nonintegrable system will tend to 
lead to a bigger spread of frequencies entering the expression for 
the overlap $A(t)$, and so to a greater degree of damping in 
$P(t)$ \cite{baldwin}.

We can also relate the behavior of $P(t)$ to 
conserved quantities within the problem.
For free electrons $H = H_{0 \parallel} $ and, since $[H_{0 \parallel}, 
H_{0 \perp}] = 0$ we may use the Baker-Hausdorf identity
$e^{A+B} = e^A e^B e^{-\frac{1}{2}[A,B]}$ (where $[A, [A,B]] = [B, [A,B]] 
=0$)
to combine the terms in $A(t)$, and show as above that

\begin {eqnarray}
P(t) = |<\psi_0|e^{i H_{0 \perp}t}|\psi_0 >|^2
=[\cos (t_{\perp}t)]^{2 \Delta N}.
\end {eqnarray}
Making a change to binding/antibinding coordinates
\begin{eqnarray}
c^{(1)\dagger}_{\sigma}(k) &=& 
 \frac{1}{\sqrt{2}} ( b^{\dagger}_{\sigma}(k) + 
a^{\dagger}_{\sigma}(k))\nonumber\\
c^{(2)\dagger}_{\sigma}(k) &=&
 \frac{1}{\sqrt{2}} ( b^{\dagger}_{\sigma}(k) - a^{\dagger}_{\sigma}(k) )
\end{eqnarray}

\noindent
we see that the condition for {\it complete} coherence is that the 
Hamiltonian $H$ commute with
\begin{eqnarray}
H_{0 \perp} = - t_{\perp} \sum_{k, \sigma} 
[b^{\dagger}_{\sigma}(k) b_{\sigma}(k) - a^{\dagger}_{\sigma}(k)a_{\sigma}(k)],
\end{eqnarray}
{\it i.e.\/} that the difference in the number of particles in the binding and 
antibinding bands be a good quantum number of the system; if the original 
Hamiltonian may be diagonalized in the basis of eigenstates of the 
perturbation then the original state will evolve completely coherently. The 
introduction of interaction can quickly be seen to 
spoil this property of the Hamiltonian.  For example, for
the two chain Hubbard Model
\begin{eqnarray}
H &=& H_1 + H_2 + H_{0 \perp}\nonumber\\
H_1 &=& -t_{\parallel} \sum_{i, \sigma} 
              \psi^{\dagger (1)}_{i+1 \sigma} \psi^{(1)}_{i \sigma} 
              \psi^{\dagger (1)}_{i \sigma} \psi^{(1)}_{i+1 \sigma}
    + U \sum_{i, \sigma} n^{(1)}_{i, \sigma} n^{(1)}_{i -\sigma}\nonumber\\
H_{0 \perp} &=& -t_{\perp} \sum_{i \sigma}
             \psi^{\dagger  (1)}_{i \sigma} \psi^{(2)}_{i \sigma}    
           + \psi^{\dagger  (2)}_{i \sigma} \psi^{(1)}_{i \sigma},
\end{eqnarray}

\noindent
giving
\begin{eqnarray} 
[H_{0\perp}, H] &=& -t_{\perp}U \sum_{i, \sigma}
    n^{(1)}_{i \sigma}[  \psi^{\dagger  (1)}_{i -\sigma} \psi^{(2)}_{i -\sigma}    
                       + \psi^{\dagger  (2)}_{i -\sigma} \psi^{(1)}_{i -\sigma} 
]
 +  n^{(2)}_{i \sigma}[  \psi^{\dagger  (2)}_{i -\sigma} \psi^{(1)}_{i -\sigma}    
                       + \psi^{\dagger  (1)}_{i -\sigma} \psi^{(2)}_{i -\sigma} 
] .
\label{eq:commut}
\end{eqnarray} 

Th quantity  $[H_{\perp}, H]$ does not commute with either part of $H$,  
on account of the terms of the form 
$ \psi^{\dagger (b)}_{i, \sigma} \psi^{(a)}_{i, \sigma}
  \psi^{\dagger (a)}_{i, -\sigma} \psi^{(b)}_{i, -\sigma} $
in the interaction, so complete coherence is lost.  At a microscopic level, 
this need not mean that interchain hopping is entirely incoherent, only 
that $N^{(b)} - N^{(a)}$ is not a good quantum number for the system and 
that the interchain hopping has some incoherent part.  We 
might expect the orthogonality of the interacting and noninteracting 
ground states of the 1D Fermi gas (measured by the parameter $\alpha$) 
to render interchain hopping highly incoherent, and so frustrate the formation
of band structure in the interchain direction.  In fact the issue is more subtle
than this; the effect of the
commutator (\ref{eq:commut}) could be cancelled by the introduction of
an additional interaction term 
$U \sum_{i, \sigma} n^{(2)}_{i, \sigma} n^{(1)}_{i -\sigma}$.
This {\it will not} restore fermionic quasiparticle character to the electronic
states of the individual chains in the absence of $H_{0 \perp}$, but {\it will} 
make possible the macroscopically coherent evolution of the specially prepared
state described above.  Thus we must either reject our macroscopic 
criterion for coherence described above, or accept that hopping between
non-Fermi liquids can be entirely coherent.  We draw the latter conclusion and 
illustrate this point in more detail for
a continuum model in the following section.

\section{A Soluble Two Chain Model with Anomalous Dimension   
              and Coherent Interchain Hopping}  

We take as a general Hamiltonian for the coupled chain problem :

\begin{eqnarray}
\label{H}
H &=& H_0 + V_{\parallel} + V_{\perp}.
\end{eqnarray}

\noindent
Here,

\begin{eqnarray}
\label{H_0} 
H_0 &=& \sum_{k, \sigma, m} \epsilon_L(k) 
           [ c^{(1) \dagger}_{L\sigma}(k) c^{(1)}_{L\sigma}(k)
           + c^{(2) \dagger}_{L\sigma(k)} c^{(2)}_{L\sigma}(k) ]\nonumber\\
       &-& t_{\perp} [ c^{(1) \dagger}_{L\sigma}(k) c^{(2)}_{L\sigma}(k)
           + c^{(2) \dagger}_{L\sigma}(k) c^{(1)}_{L\sigma}(k) ] 
                       + [ L \rightarrow R ], 
\end{eqnarray} 

\begin{eqnarray}
V_{\parallel} &=& \frac{1}{2L} \sum_{q>0, \sigma, \sigma^{\prime}}
                 g_{4\parallel} [\rho^{(1)}_{L\sigma}(-q) 
                                 \rho^{(1)}_{L\sigma^{\prime}}(q)
                              +  \rho^{(2)}_{L\sigma}(-q) 
                                 \rho^{(2)}_{L\sigma^{\prime}}(q)]\nonumber \\
               &+& g_{2\parallel} [\rho^{(1)}_{L\sigma}(-q)
                                 \rho^{(1)}_{R\sigma^{\prime}}(q)  
                               + \rho^{(2)}_{L\sigma}(-q)
                                 \rho^{(2)}_{R\sigma^{\prime}}(q)] \nonumber \\
               &+& g_{4\parallel} [(N^{(1)}_{L\sigma}
                                  - N^{(1)}_{L0\sigma^{\prime}})^2
                              +  (N^{(2)}_{L\sigma}
                                 - N^{(2)}_{L0\sigma^{\prime}})^2]\nonumber\\
               &+& g_{2\parallel} [(N^{(1)}_{L\sigma} - N^{(1)}_{L0\sigma})
                                   (N^{(1)}_{R\sigma^{\prime}}
                                    - N^{(1)}_{R0\sigma^{\prime}})  
                               + (N^{(2)}_{L\sigma} - N^{(2)}_{L0\sigma})
                                (N^{(2)}_{R\sigma^{\prime}}
                                 - N^{(2)}_{R0\sigma^{\prime}}]\nonumber\\
               &+& [ L \rightarrow R , q \rightarrow -q ] 
\end{eqnarray}

\begin{eqnarray}
V_{\perp} &=& \frac{1}{2L} \sum_{q>0, \sigma, \sigma^{\prime}}
                 g_{4\perp} [\rho^{(1)}_{L\sigma}(-q) 
                             \rho^{(2)}_{L\sigma^{\prime}}(q)
                          +  \rho^{(2)}_{L\sigma}(-q) 
                             \rho^{(1)}_{L\sigma^{\prime}}(q)] \nonumber \\
               &+& g_{2\perp} [\rho^{(1)}_{L\sigma} 
                             \rho^{(2)}_{R\sigma^{\prime}}(q)  
                           + \rho^{(2)}_{L\sigma}(-q)
                             \rho^{(1)}_{R\sigma^{\prime}}(q)] \nonumber \\
               &+& g_{4\perp} [(N^{(1)}_{L\sigma}
                                  - N^{(1)}_{L0\sigma^{\prime}})^2
                              +  (N^{(2)}_{L\sigma}
                                 - N^{(2)}_{L0\sigma^{\prime}})^2]\nonumber\\
               &+& g_{2\perp} [(N^{(1)}_{L\sigma} - N^{(1)}_{L0\sigma})
                                   (N^{(1)}_{R\sigma^{\prime}}
                                    - N^{(1)}_{R0\sigma^{\prime}})  
                               + (N^{(2)}_{L\sigma} - N^{(2)}_{L0\sigma})
                                (N^{(2)}_{R\sigma^{\prime}}
                                 N^{(2)}_{R0\sigma^{\prime}}]\nonumber\\
               &+& [ L \rightarrow R , q \rightarrow -q ]
\end{eqnarray}

\begin{eqnarray}
\rho^{(i)}_{m \sigma}(q)  &=& \sum_{p \sigma } 
        c^{(i) \dagger}_{m \sigma}(p+q) c^{(i)}_{m\sigma} (p)
\end{eqnarray}

\begin{eqnarray}
N^{(i)}_{m \sigma} &=& \sum_{p \sigma } 
        c^{(i) \dagger}_{m \sigma}(p) c^{(i)}_{m\sigma} (p)
\end{eqnarray}

\noindent
where $c^{(i)\dagger}_{m\sigma}(k)$ is the fermion 
creation operator with spin index $\sigma = \pm 1$, on the left (right) 
moving 
branches ($m = L/R$) of the chain with index $i=1/2$, 
and $\epsilon_{R/L}(k) = v_f(\pm k-k_f)$ (see Figure \ref{dispersion} 
for clarification of labels).  
Neglecting all branch mixing `backscattering' events, all four 
interactions in Figure \ref{interaction} have been expressed above in terms of electronic 
densities $\rho^{(i)}_{m\sigma}(q)$.  The 
$q=0$ component of the interaction terms, which contains information about 
chemical potential in the form of a bare charge $N^{(i)}_{m0\sigma}$, 
has been explicitly included. This will be seen to be important 
in what follows.  We have also made the nonrestrictive 
assumption of a spin independent interaction so that 
$g^{\uparrow \uparrow} = g^{\uparrow \downarrow} =g$.  
This may easily be relaxed.

To illustrate more clearly the fact that the spectral properties of an 
isolated chain are not always an accurate guide to the coherence of 
interchain hopping in a two chain model, we will consider below a special 
soluble point of the interacting problem.  In the interests of clarity we will
suppress spin dependence. We will look at the issue
of spin-charge separation later.  (In fact the generalization of the model to 
include spin is straightforward, and does not affect any of the arguments of 
this section, which relate chiefly to the anomalous exponent $\alpha$.)

If we rewrite the momentum-space Hamiltonian in terms of collective coordinates 
\begin{eqnarray}
\alpha^{\dagger}(q) &=& \frac{1}{\sqrt{2}} \frac{2\pi}{qL} 
                        [ \rho^{(b)}_R (q) + \rho^{(a)}_R (q) ]\nonumber\\
\beta^{\dagger}(q) &=& \frac{1}{\sqrt{2}} \frac{2\pi}{qL} 
                        [ \rho^{(b)}_L (-q) + \rho^{(a)}_L (-q) ]\nonumber\\
\gamma^{\dagger}(q) &=& \frac{1}{\sqrt{2}} \frac{2\pi}{qL} 
                        [ \rho^{(b)}_R (q) - \rho^{(a)}_R (q) ]\nonumber\\
\delta^{\dagger}(q) &=& \frac{1}{\sqrt{2}} \frac{2\pi}{qL} 
                        [ \rho^{(b)}_L (q) - \rho^{(a)}_L (q) ]       \\
\label{eq:albegade}
\end{eqnarray}                        
with $[\alpha(q), \alpha^{\dagger}(q)] = \delta_{q q^{\prime}}$, 
$[\alpha, \beta] =0$ etc., we find that the kinetic energy and interaction 
parts of the Hamiltonian take on a particularly simple form, provided that 
we impose the restriction $g_{4\parallel }=g_{4\perp }=g_4$, 
$g_{2\parallel}=g_{2\perp}=g_2$.  Then :
\begin{eqnarray}
H_0 &=& \sum_{k \sigma} v_f q [\alpha^{\dagger}(q) \alpha(q) 
                              + \beta^{\dagger}(q) \beta(q)
                              + \gamma^{\dagger}(q) \gamma(q)
                             + \delta^{\dagger}(q) \delta(q) ]\nonumber\\
    &+& \frac{\pi v_f}{L} [ (N_{\alpha} - N^0_{\alpha})^2 
                          + (N_{\beta} - N^0_{\beta})^2 
                          + (N_{\gamma} - N^0_{\gamma})^2
                          + (N_{\delta} - N^0_{\delta})^2 ],
\label{eq:H0}
\end{eqnarray} 
and
\begin{eqnarray}
V &=& \sum_{q >0} \frac{g_4 q}{L} [ \alpha^{\dagger}(q) \alpha(q) 
                             + \beta^{\dagger}(q) \beta(q) ] \nonumber\\
  &+& \sum_{q >0} \frac{g_2 q}{L} [ \alpha^{\dagger}(q) \beta^{\dagger}(-q) 
                             + \alpha(-q) \beta(q) ] \nonumber\\
  &+& \frac{g_4 q}{L} [ (N_{\alpha} - N_{\alpha}^0)^2 
                             + (N_{\beta} - N_{\beta}^0)^2 ] \nonumber\\
  &+& \frac{2 g_2 q}{L} [ (N_{\alpha} - N_{\alpha}^0) 
                                                (N_{\beta} - N_{\beta}^0)],
\end{eqnarray}
with $N^0_{\alpha} = N^0_{\beta} = \sqrt{2}N^0$ and 
$N^0_{\gamma} = N^0_{\delta} = \frac{t_{\perp}L}{\sqrt{2} \pi v_f}$.

Physically this choice of interaction
corresponds to having equally strong interchain and 
intrachain interaction, at least for the small $q$ regime of the 
bosonization.  While this condition is unlikely to be met in the 
majority of quasi-one dimensional conductors, we  note in passing 
that a number of `ladder' compounds, with 
nearly isotropic coupling between well spaced pairs of one-dimensional 
chains 
have recently been synthesized \cite{rice}.

The Hamiltonian for the spinless two chain system above is then quadratic, 
and describes two independent one dimensional systems.
It may be solved directly by canonical transformation.  We can therefore 
proceed 
to calculate the correlation functions for the particles in binding and 
antibinding bands, using the methods standard in the bosonization 
literature.
The correct non-interacting (retarded) Greens function can be correctly 
recovered 
from the bosonic expression for $H_0$ using
\cite{Rev_Boson} to give 
\begin{eqnarray}
< \psi^{\dagger}_{1R}(x,t) \psi_{1R}(0,0) >_0  & = &
   \frac{i}{2\pi} \frac{\exp{-ik_fx}}{x-v_ft+i\epsilon} 
   \cos [\frac{t_{\perp}x}{v_f}],
\end{eqnarray}
demonstrating the continuity between the collective bosonic and 
single-particle fermionic descriptions at the noninteracting point.
The effect of interchain hopping on free electrons is to 
introduce a cosine modulation in the single particle correlation 
function on a single chain reflecting the Fermi surface shifts associated 
with the formation of binding and antibinding states.  This vanishes 
continuously 
as $t_{\perp} \rightarrow 0$.  We may find the correlation function for 
our 
spinless interacting model by direct analogy with careful treatment of the 
one chain spinful case \cite{voit}
\begin{eqnarray}
 < \psi^{\dagger}_{1R}(x,t) \psi_{1R}(0,0) >  & = &  
\frac{i}{2\pi} \exp{(-ik_fx)} 
  \frac{\Lambda + i(v_f t - x)}{\epsilon + i(v_f t - x)} 
    \nonumber \\
  & & \mbox{ } 
  \frac{1}{ [x-v_{\alpha}t+i\epsilon]^{\frac{1}{2}}
               [x-v_f t+i\epsilon]^{\frac{1}{2}} }
     \left( \frac{\Lambda^2}{ (\Lambda + i v_{\alpha} t)^2 + x^2 } 
     \right)^{\alpha/2}  
       \cos{[\frac{t_{\perp}x}{v_f}]},
\end{eqnarray}
where $\alpha = [1-(g_2/(g_4 + \pi v_F))^2]^{-1/2} - 1$
 and $v_{\alpha} =  [(v_F + g_4/\pi )^2 - (g_2/\pi )^2]^{1/2}$.  
 $\Lambda$ is a length scale set by the 
range of interaction.  The effects of interaction on the model are 
apparent in  the branch cut (analogous to ``spin/charge'' separation 
$v_{\alpha} \ne v_f$), and in 
the anomalous scaling $G(sx, st) = s^{-1-\alpha} G(x,t)$.  The eigenstates 
of 
this spinless coupled two chain model are clearly bosonic, 
and it is possible to show
explicitly that the interacting ground state is orthogonal to the
noninteracting ground state.  We wish to emphasize 
that the same results are obtained in the limit $t_{\perp} \rightarrow 0$ 
as in the absence of $t_{\perp}$, and to remark that this model may be 
mapped onto the case of a single chain of spinful electrons in a magnetic field.

We also learn from this correlation function that the Fermi surface shifts 
{\it 
have} survived the introduction of interaction, despite the consequent 
Luttinger Liquid character of motion along the chains.   The momentum 
distribution $n(k)$ will have power law behavior $n(k-k_f) \sim 
(k-k_f)^{\alpha}$ 
in 
both the binding and antibinding bands, but with different values of $k_f$ 
in 
each band.

This is strongly suggestive of coherent electronic motion between the 
chains, and in fact for our model with the special choice 
$g_{2\perp} = g_{2\parallel} = g_2$,
$g_{4\perp} = g_{4\parallel} = g_4$, 
\begin{eqnarray}
[H_{0\perp}, V_{\parallel}] =  - [H_{0\perp}, V_{\perp}],  
\end{eqnarray}
\noindent
and the quantity $P(t)$ (\ref{eq:Poft}) behaves 
exactly as in a free electron gas.  This is despite the 
fact that the 
single particle correlation  function {\it does not} possess a pole
at the Fermi momentum, and that 
the interacting groundstate is orthogonal to the non interacting ground state.  
We 
understand from the commutator above that the source of any incoherence of 
hopping 
between chains in 
a more general two chain model are those non-branch conserving terms in the 
Hamiltonian which render 
the model insoluble by bosonization, and which give rise to the rich variety
of dominant fluctuations found by scaling arguments.

An alternative physically motivated test for the coherence of hopping 
between 
the chains is provided by the interchain polarization response function, 
whose 
associated spectral function is given by
\begin{eqnarray}
\rho_{\Delta N} (w) = \int_{0}^{\infty} dt e^{-i\omega t} 
\Im { <[\Delta N(t), \Delta N(0)]> }
\end{eqnarray}
where the operator $\Delta N = N_1 - N_2$ is the difference between the 
number 
operators for a given branch of the two chains.  For free electrons 
with linear dispersion this may 
easily be shown to 
be monochromatic---the only frequency entering the response is the 
binding/antibinding  energy gap $2 t_{\perp}$.   We may recover the same 
result 
from the bosonic form of $H_0$ by calculating $[H_0, [H_0, \Delta N]]$ to 
show
\begin{eqnarray}
\frac{ {\partial}^2 \Delta N }{ \partial t^2 } &=& - (2t_{\perp})^2 \Delta N
\end{eqnarray}
reflecting the fact that $\Delta N $ is {\it not} a good quantum number in 
the presence of hopping.  This commutator would vanish, in the 
both interacting and the free electron case, were the chemical potential terms in the 
Hamiltonian to be neglected.  The finite expectation value of 
$(\Delta N)^2 \sim 2t_{\perp}$ in the noninteracting ground state may then 
be understood in terms of the zero point motion of a simple harmonic 
oscillator 
with $\Delta N$ as its coordinate, and canonical momentum proportional to 
the current operator 
$c^{\dagger}_1 c_2 - c^{\dagger}_2 c_1$.  The raising operator for this 
oscillator is 
in turn proportional to $N_{\gamma}$ (for the right moving branch,
see \ref{eq:albegade}), with 
$<N_b + N_a> = <N_1 + N_2>$ a constant of motion with or without 
$t_{\perp}=0$.  The inclusion of our special interaction does not change 
this result; again we 
find a monochromatic polarization response function and 
$\Delta N^2 \sim 2t_{\perp}$, confirming the coherent nature of the 
hopping.  

We see that it is possible to propose a model for a two chain system which is 
a Luttinger Liquid in the chain direction 
{\it for any finite} $g_2$, and to show directly from the
Hamiltonian, from the correlation functions, or from the response of the 
system to interchain polarization, that it displays fully coherent hopping 
between the chains, with 
associated shifts in the ``Fermi-surface" {\it for any finite} $t_{\perp}$.  
These two 
properties are thus found in this special case to be entirely independent 
of one another; anomalous scaling in the chain direction coexists with 
coherent hopping in the interchain direction.  The intuitive idea that the 
spectral properties of the model for $t_{\perp} = 0$ determine the possibility 
of confinement at finite $t_{\perp}$ is seen in this instance to fail 
completely. 

We may generalize those arguments which relate to the 
commutation relations of the hopping term to N chains simply by 
including a nearest neighbor interchain interaction 
between each pair of chains to cancel the undesired commutator
$[V_{\parallel}, H_{0\perp}]$.  However the generalization of 
the bosonization scheme to the evaluation of correlators for the 
N-chain case is not quite so straightforward.  Thus while it 
seems to be possible to
achieve coherent interchain transport in a two chain model 
without the formation of a pole in the correlation function, it 
is not clear that we may make the same statement about an N chain 
model.  Our two chain model is also insensitive to the size of 
$t_{\perp}$, but should not be considered physical for any value 
of $t_{\perp}$ bigger than a scale set by the 
curvature of the real free fermion dispersion and the interaction range 
$\Lambda$).

\section{Coherence of Hopping and Spin Charge Separation}

Spin charge separation is perhaps the best known of the unusual 
properties of the one dimensional interacting  electron gas.   In the 
Luttinger model (for which the 
Luttinger Liquid ground state is an exact solution), it has its 
origin in the fact that the elementary excitations of the gas are not 
fermionic quasiparticles, but rather bosonic collective modes.  In the 
presence of a spin-dependent interaction (such as an exchange potential), 
the modes associated with spin and charge degrees of freedom need not 
disperse with the same velocity.

A hopping term of the form of $H_{0 \perp}$,
however, must transfer a `real' electron, with a corresponding 
non-trivial adjustment in the spin and  charge excitations on each 
chain.  Naively we might expect this to frustrate hopping between chains, 
and so protect (at least partially) the one-dimensional interacting 
Fermi gas from the destabilizing effects of higher dimension.   Even if 
spin charge separation alone is not enough to generate confinement of the 
electrons on a single chain, it might be expected to reduce the coherence 
of hopping between chains and, by reducing the 
`phase memory' of the transferred electron, to
hinder the formation of band structure perpendicular to the 
chains.  In this section, we will consider a soluble one branch, spinful, 
two-chain model with inter- and intra-chain $g_4$ interaction, and 
assess the effect of the spin-charge separation which it exhibits on the
coherence of hopping between chains.

We may solve the one branch model of an electron 
gas with linear dispersion and forward scattering interactions 
by bosonization :
\begin{eqnarray}
H_1 &=& \sum_{k \sigma} v_f  (k-k_f) 
              c^{\dagger (1)}_{\sigma}(k) c^{(1)}_{\sigma}(k) 
           + \frac{1}{2L} \sum_{q \sigma \sigma^{\prime}}
            (g_{4 \parallel}^{\uparrow \uparrow} 
                  \delta_{ \sigma \sigma^{\prime}}
            +g_{4 \parallel}^{\uparrow \downarrow} 
                  \delta_{ \sigma -\sigma^{\prime}})
            \rho^{(1)}_{\sigma} (q) \rho^{(1)}_{\sigma^{\prime}} (-q)
\end{eqnarray}
where $ \rho^{(1)}_{\sigma} (q)$ is the Fourier transform of the density 
operator for electrons with spin index $\sigma$ on the chain with index 
$1$.  
In fact it is possible also to solve exactly the two chain problem 
\begin{eqnarray}
H &=& H_1 + H_2 + H_{0 \perp}
\end{eqnarray}
\begin{eqnarray}
H_{0 \perp} &=& -t_{\perp} \sum_{k \sigma}
         c^{\dagger  (1)}_{\sigma}(k) c^{(2)}_{\sigma}(k)    
       + c^{\dagger  (2)}_{\sigma}(k) c^{(1)}_{\sigma}(k)
\end{eqnarray}
by expressing $H_{0 \perp}$ in terms of the bosonic representation of the 
electronic field operators on each chain, as illustrated by 
Fabrizio and Parola \cite{fab1,fab2}.
The Hamiltonian can then be expressed in a basis comprising four 
branches of spinless fermions with linear dispersion
\begin{eqnarray}
H_1 + H_2 &\rightarrow& \sum_{k>0} 
                   u_{\rho}k [ a^{\dagger}_{\rho}(k) a_{\rho} (k)
                             + b^{\dagger}_{\rho}(k) b_{\rho} (k) ]
                 + u_{\sigma}k [ a^{\dagger}_{\sigma}(k) a_{\sigma} (k)
                               + b^{\dagger}_{\sigma}(k) b_{\sigma} (k) ],
\end{eqnarray}
\begin{eqnarray}
H_{0 \perp} &\rightarrow& -2t_{\perp} \sum_{k>0} 
                   b^{\dagger}_{\rho}(k) b_{\sigma} (k) 
                 + b^{\dagger}_{\sigma}(k) b_{\rho} (k) .
\end{eqnarray}
where,
\begin{eqnarray}
u_{\rho} &=& v_f + \frac{( g_{4 \parallel}^{\uparrow \uparrow} 
                         + g_{4 \parallel}^{\uparrow \downarrow} )}
                         {2\pi}\nonumber\\
u_{\sigma} &=& v_f + \frac{( g_{4 \parallel}^{\uparrow \uparrow} 
                           - g_{4 \parallel}^{\uparrow \downarrow} )} 
                         {2\pi}.
\end{eqnarray}
This may be diagonalized by canonical transformation to yield four 
branchs of free, spinless, fermions with dispersion

\begin{eqnarray}
\epsilon_{\rho} (k) &=& u_{\rho}k \nonumber\\
\epsilon_{\sigma} (k) &=& u_{\sigma}k \nonumber\\
\epsilon_{\pm} (k) &=& \frac{1}{2} (u_{\rho} + u_{\sigma}) k
                         \pm \sqrt{[\frac{1}{2}(u_{\rho}-u_{\sigma})^2
                                         + 4 t_{\perp}^2},
\end{eqnarray}

\noindent
and the ground state found by filling the fermion branch with dispersion 
$\epsilon_-(k)$ up to the chemical potential $\mu = 0$.  The expectation 
value of the number operator $N^{(b)} - N^{(a)}$ can be calculated and is
strongly suggestive of the formation of `band structure' in the interchain 
direction :
\begin{eqnarray}
\frac{< N^{(b)} - N^{(a)} >}{L} &=&  
    \frac{1}{u_{\rho} - u_{\sigma}} \ln(\frac{u_{\rho}}{u_{\sigma}})
    \frac{4t_{\perp}}{2\pi}.
\label{eq:NaNb}
\end{eqnarray}

In fact the commutator 
$[H_1 + H_2, H_{\perp}]$ {\it does not} vanish, so $N^{(b)} - N^{(a)}$ 
is {\it not} a good quantum number for the system, but it is clear 
that the splitting of energy levels due to $t_{\perp}$ is only 
partially suppressed.  Those terms in the Hamiltonian which 
do not commute with $N^{(b)} - N^{(a)}$ give rise to the prefactor
\begin{eqnarray}
\frac{1}{u_{\rho} - u_{\sigma}} \ln(\frac{u_{\rho}}{u_{\sigma}}) \leq 1
\label{eq:reduc}
\end{eqnarray}
which would be unity in the free electron case.  It is 
a consequence of the curvature in $\epsilon_{-} (k)$, which depends on both  
$t_{\perp}$ and $u_{\rho} - u_{\sigma}$, and indicates a suppression of 
dispersion in the interchain direction in the presence of spin charge 
separation $u_{\rho} \neq u_{\sigma}$.  As such it may be taken as a 
measure of the incoherence of hopping between chains. 

It is not difficult to generalize the model of Fabrizio and Parola to 
include interchain interaction; the structure of the Hamiltonian is 
altered only in that the $a(k)$ and $b(k)$ fermions acquire different velocities 
\begin{eqnarray}
H_1 + H_2 &\rightarrow& \sum_{k>0} 
                 [ u_{\rho a}k a^{\dagger}_{\rho}(k) a_{\rho} (k)
                 + u_{\rho b}k b^{\dagger}_{\rho}(k) b_{\rho} (k) 
                 + u_{\sigma a}k a^{\dagger}_{\sigma}(k) a_{\sigma} (k)
                 + u_{\sigma b}k b^{\dagger}_{\sigma}(k) b_{\sigma} (k) ],                  
\end{eqnarray}
where,
\begin{eqnarray}
u_{\rho a} &=& v_f + \frac{( g_{4 \parallel}^{\uparrow \uparrow} 
                           + g_{4 \parallel}^{\uparrow \downarrow} 
                           + g_{4 \perp}^{\uparrow \uparrow} 
                           + g_{4 \perp}^{\uparrow \downarrow})}
                                 {2\pi}\nonumber\\
u_{\rho b} &=& v_f + \frac{( g_{4 \parallel}^{\uparrow \uparrow} 
                           + g_{4 \parallel}^{\uparrow \downarrow} 
                           - g_{4 \perp}^{\uparrow \uparrow} 
                           - g_{4 \perp}^{\uparrow \downarrow})}
                                 {2\pi}\nonumber\\
u_{\sigma a} &=& v_f + \frac{( g_{4 \parallel}^{\uparrow \uparrow} 
                           - g_{4 \parallel}^{\uparrow \downarrow} 
                           + g_{4 \perp}^{\uparrow \uparrow} 
                           + g_{4 \perp}^{\uparrow \downarrow})}
                                 {2\pi}\nonumber\\
u_{\sigma b} &=& v_f + \frac{( g_{4 \parallel}^{\uparrow \uparrow} 
                           - g_{4 \parallel}^{\uparrow \downarrow} 
                           - g_{4 \perp}^{\uparrow \uparrow} 
                           + g_{4 \perp}^{\uparrow \downarrow})}
                                 {2\pi}.                                 
\end{eqnarray}

The dispersion of the four fermionic branches after diagonalization are 
thus modified to give
\begin{eqnarray}
\epsilon_{\rho} (k) &=& u_{\rho a}k \nonumber\\
\epsilon_{\sigma} (k) &=& u_{\sigma a}k \nonumber\\
\epsilon_{\pm} (k) &=& \frac{1}{2} (u_{\rho b} + u_{\sigma b}) k
                         \pm \sqrt{[\frac{1}{2}(u_{\rho b}-u_{\sigma b})^2
                                         + 4 t_{\perp}^2}
\end{eqnarray}
The ground state expectation value of $N^{(b)} - N^{(a)}$ is then 
given by the same
formula as before (\ref{eq:NaNb}) but with $u_\rho$ and 
$u_\sigma$ replaced by
$u_{\rho b}$ and $u_{\sigma b}$.
There is full coherence (the prefactor (\ref{eq:reduc}) equal to one)
when the dispersion $\epsilon_\pm(k)$ is linear in $k$. This happens
not just when all $g_4$'s
are equal to zero (the non-interacting case)
but also when $g^{\uparrow \downarrow}_{4 \parallel} = 
g^{\uparrow \downarrow}_{4 \perp}$. 
In either case $[H_1 + H_2, H_{\perp }] = 0$.
In the special case of equal inter and intrachain interactions,
we may solve the model directly by bosonization, obtaining the same 
ground state and correlators as are found by refermionization :
\begin{eqnarray}
< \psi^{\dagger}_{0}(x,t) \psi_{0}(0,0) > &\sim&
 e^{i(k_f + \Delta k_f)x} (x - v_f t + i\epsilon)^{-1/2}
                          (x - u_{\rho b} t + i\epsilon)^{-1/4} 
                          (x - u_{\sigma b} t + i\epsilon)^{-1/4},  
\end{eqnarray}  
where $\Delta k_f = \frac{t_{\perp}}{v_f}$.

In the general case, the prefactor (\ref{eq:reduc}) is less than
one and depends on the relative strength of the interchain
and interchain interactions.
This result is consistent with the arguments of 
CSA \cite{clarke} who also consider the model studied by 
Fabrizio and Parola. They find evidence of 
macroscopic coherence for times much less than 
$[\frac{2\pi \Delta N}{L} (u_{\rho} - u_{\sigma})]^{-1}$ \cite{shann}. 

\section{conclusions}

We have shown that it is possible for a two-chain system 
to show both coherent hopping between chains and the anomalous
intrachain scaling characteristic of a Luttinger liquid. 
For the two chain Luttinger model, which we can solve exactly
for a special choice of interchain interaction, we do not find any 
evidence for the formation of a pole at the Fermi surface.
In general, we find that the degree of coherence observed
in the interchain hopping depends
rather sensitively on the precise details of the interactions
both on the chains and between the chains and may not always
be estimated from the spectral properties of the individual
chains alone. 

A simple one-branch model exhibiting spin-charge separation is found to be
neither completely coherent nor completely incoherent in its macroscopic 
evolution, and we draw the conclusion,
in line with other authors, that spin charge separation alone may suppress, 
but is not sufficient to
prevent the formation of band structure in the hopping direction.  Again, 
for a special choice of interchain interaction it is possible to arrive at 
a model which displays both spin charge separation and completely coherent 
hopping between chains.

Although the restriction on interaction imposed by the criterion for 
coherence we have used is rather severe, and gives evidence of 
the stability of a Luttinger Liquid ground state only on the line 
(plane, in spinful case) $g_{\perp} = g_{\parallel}$, the states 
found away from this line (plane) are known from scaling arguments
to be ordered.  We speculate that at temperatures greater than a scale 
set by the gap of each ordered phase, the behavior of the system
will be controlled by this $g_{\perp} = g_{\parallel}$ (quantum
critical) line.

Nic Shannon wishes to acknowledge helpful conversations with 
David Clarke, Andrey Chubakov, George Rowlands, Edmound Orignac, 
Carol Hayward, Sylvan Capponi and
Thanassis Yannacopoulis, and also an HCM grant for attendance at 
the ISI Workshop 
on Strong Correlation in Low Dimensional Systems held in May 1996.

\begin{figure}

\caption{Dispersion relations for the four branches of 
the two chain Luttinger model, showing interchain hopping 
$t_{\perp}$.}

\label{dispersion}

\end{figure}

\begin{figure}

\caption{Examples of chain and branch indices for 
the four interaction terms kept in the bosonic 
treatment of the two chain problem.}

\label{interaction}

\end{figure}
                                      

\begin{references}
\bibitem{correction} This is in fact not quite correct.  If we explicitly
                     include the shift in the value of $k_f$ required to 
                     restore the chemical potential to zero on the chain 
                     with additional charge $\Delta N$, we find eigenvalues
                     $\epsilon (k) = v_f(k-k_f) + \frac{\Delta \mu}{2}
                     \pm \sqrt{t^2_{\perp} + {\Delta \mu}^2}$ where
                     $\Delta \mu = \frac{2 \pi v_f}{L} \Delta N$, corresponding
                     to a slightly different rotation of coordinates from the
                     binding/antibinding transformation we shall use, and 
                     modifying the frequency of oscillation between levels from
                     $t_{\perp}$ to $t_{\perp}^* = 
                      \sqrt{t^2_{\perp} + {\Delta \mu}^2}$ .  This is a small 
                     quantitative effect, and modifies our results only in the 
                     correction $t_{\perp} \rightarrow t_{\perp}^*$,
                     so we shall neglect it in what follows.
\end{references}
\end{document}